\documentclass[aps,prd,showpacs,amsfonts,superscriptaddress]{revtex4}
\usepackage{latexsym}
\usepackage{amsmath}
\def\cM{{\cal M}}

\def\RE{{\rm I\!R}}
\def\NA{{\rm I\!N}}
\def\lo{{\!{}_{\vec{\lambda}=\vec{0}}}}
\def\nn{\nonumber}
\def\NN{\mbox{\tiny N}}
\def\MM{\mbox{\tiny M}}
\def\PP{\mbox{\tiny P}}
\def\QQ{\mbox{\tiny Q}}
\def\RR{\mbox{\tiny R}}
\def\e{\mbox{e}}

\def\vp{{}^\varphi\!}
\def\ps{{}^\psi\!}

\def\CMP{Commun. Math. Phys.~}
\def\CQG{Class. Quant. Grav.~}
\def\PRD{Phys. Rev. D~}
\def\PRL{Phys. Rev. Lett.~}

\begin{document}
\title{Nonlinear N-parameter spacetime perturbations: Gauge transformations}
\author{Carlos F. Sopuerta}
\affiliation{Institute of Cosmology and Gravitation,
University of Portsmouth, Portsmouth PO1 2EG, Britain}
\affiliation{Institute for Gravitational Physics and Geometry and Center for
Gravitational Wave Physics, Penn State University, University Park,
Pennsylvania 16802, USA}
\author{Marco Bruni}
\affiliation{Institute of Cosmology and Gravitation,
University of Portsmouth, Portsmouth PO1 2EG, Britain}
\author{Leonardo Gualtieri}
\affiliation{Dipartimento di Fisica ``G. Marconi'', Universit\`a di Roma
``La Sapienza'' and Sezione INFN ROMA 1, piazzale Aldo Moro 2, I--00185 Roma,
Italy}

%%%%%%%%%%%%%%%%%%%%%%%%%%%   DATE  %%%%%%%%%%%%%%%%%%%%%%%%%%%%%%%%%%

\date{\today}

%%%%%%%%%%%%%%%%%%%%  ABSTRACT  %%%%%%%%%%%%%%%%%%%%%%%%%%%%%%%%%%%%%%

\begin{abstract}
We introduce N-parameter perturbation theory as a new tool for the study of non-linear
relativistic phenomena.  The main ingredient in this formulation is the use of the
Baker-Campbell-Hausdorff formula.  The associated machinery allows us to prove the main
results concerning the consistency of the scheme to any perturbative order.
Gauge transformations and conditions for gauge invariance at any required order
can then be derived from a generating exponential formula via a simple Taylor expansion.
We outline the relation between our novel formulation and previous developments.
\end{abstract}


\maketitle

%\section{Introduction}\label{intro}
In the theory of spacetime perturbations
\cite{Sa,StWa,BrMaMoSo,SoBr,BrGuSo} (see \cite{Wa,St2,Sc} for an
introduction), one usually deals with a family of spacetime models which,
in most cases, depends on a single parameter $\lambda$:
$M_{\lambda}=(\cM,\{{\cal T}_{\lambda}\})$, where $\cM$ is a
manifold that accounts for the topological and
differential properties of spacetime, and
$\{{\cal T}_{\lambda}\}$ is a set of fields on $\cM$,
representing its geometrical and physical content (this formulation
does not depend on the gravitational field equations). The
parameter $\lambda$ labels the elements
of the family and gives an indication of the ``size'' of the
perturbations, regarded as deviations of $M_{\lambda}$
from a background model $M_0$. It can either be a formal
parameter, as in cosmology \cite{BrMaMoSo,MaMoBr,BrMeTa}, in
back-reaction problems (see e.g. \cite{FlWa,Nambu}), or in the study of
quasi-normal modes of stars and black holes \cite{Ch,KoSc}, or it can have a
specific physical meaning, as in the study of  black hole mergers via
the close limit approximation,  in the analysis of quasi-normal mode excitation
by a physical source, or in the modelling of perturbations generated by the
collapse of a rotating star (see \cite{KoSc,closelimit,roma,CPM} and references
therein).

There are  however  physical applications in which it may be convenient to use
a perturbation formalism based on two \cite{BrGuSo} or more parameters.
For instance, in order to study general time-dependent perturbations of stationary
axisymmetric rotating stars \cite{rotatingpert} using a spherical background.
In this case one can separately consider the stationary axisymmetric rotational
perturbations, for example up to second order in a parameter $\Omega$, then
considering the coupling of these with the first-order time dependent ones
(see \cite{BrGuSo} for further discussion). As it should be clear from this
example, the advantage of an N-parameter Non-Linear Perturbation Theory (NLPT)
is that it allows us to make distinctions between different
types of perturbations corresponding to different parameters, so that we can study
their  coupling and some non-linear effects without having to compute the whole
set of higher-order perturbations.   Such a framework may provide flexibility by
allowing us to look at a given problem from different points of view.
It may also allow us to choose a simpler (typically more symmetrical) background
to model a given physical scenario. Given that, even in NLPT,  the differential
operators are those defined on the chosen background, this can drastically reduce
the computations, and even change the way to perform them.

The aim of the present article is to introduce a novel approach to the study
of the gauge dependence of perturbations in NLPT which (i) deals with an arbitrary
number of parameters; (ii) provides a closed formula for the action of a general gauge
transformation, valid to any order; (iii) the construction and
derivation of formulas of practical
interest is simpler than in previous frameworks \cite{SoBr,BrMaMoSo,BrGuSo}.
This new approach is mainly based on the application of  the
Baker-Campbell-Hausdorff (BCH) formula \cite{BCH}. This has been used
previously \cite{MAB}, in the context of the back-reaction problem in cosmology,
to derive second order one-parameter gauge transformations. Here we show how to
make use of the full power of the BCH formula deriving
the transformation between two given N-parameter gauges, each represented by
an N-parameter group of diffeomorphisms, at an arbitrary order. Our formalism
therefore also contains the conditions for gauge invariance for every
perturbation order in N-parameters.

We start by summarizing some relevant results regarding the mathematical
structure of the single parameter NLPT (see \cite{BrMaMoSo,SoBr}).
The Taylor expansion of a general non-linear gauge transformation can be expressed
in terms of Lie derivatives with respect to a set of vector fields which,
order by order, constitute the generators of the transformation.
A closed formula for this expansion, valid at all
orders, was found in \cite{SoBr},
using a new object dubbed {\it Knight Diffeomorphism} (KD), first defined in
\cite{BrMaMoSo} (cf.\ also \cite{FlWa,bi:GL}). The analysis in
\cite{BrMaMoSo,SoBr,BrSo} gives also the
conditions for gauge invariance at any given order, and provides the framework
for the construction of gauge-invariant formalisms \cite{CaLo}.
A similar construction has been attempted for the two-parameter
case in \cite{BrGuSo}, where the action of a general gauge transformation
on arbitrary tensor quantities was expressed in terms of the Lie
derivatives with respect to  a set of vector fields. However,
since no natural generalizations of the KD approach were
found, these expressions were derived up to fourth order in the parameters by
imposing, order by order, consistency conditions (see \cite{copione} for
a related analysis and \cite{copione2} for an application to
gravitating strings). It must be pointed out that, although that derivation is
not as elegant and compact as in \cite{SoBr,BrMaMoSo} or the one  based on the
BCH formula  presented here,
it still leads to the right formulas of practical interest, as we shall discuss.

The basic assumption for the construction of a multi-parameter
relativistic NLPT is the existence of a multi-parameter family of spacetime
models $M_{\vec{\lambda}}=(\cM,\{{\cal T}_{\vec{\lambda}}\})$, where $\cM$ denotes
the spacetime manifold
and $\{{\cal T}_{\vec{\lambda}}\}$ is a set of fields on $\cM$, describing
their geometrical and physical content, which we assume to be
analytic. These spacetime models are labeled
by a set of N parameters $\vec{\lambda}=(\lambda_1,\dots,\lambda_{\NN})$
that control the strength of the perturbations with respect to the
{\em background} spacetime model $M_{\vec{0}}$,
which describes an idealized situation.
In order to construct the physical spacetime model $M_{\vec{\lambda}}$
as a deviation from the background model $M_{\vec{0}}$, we need to establish a
correspondence between them; what in the context of relativistic perturbation
theory is called a {\em gauge choice}.
This correspondence is established, for all $\vec{\lambda}$, through the action
of a diffeomorphism of $\cM$: $\varphi_{\vec{\lambda}}: \cM\rightarrow\cM$.
The set of diffeomorphisms ${\cal G}[\varphi] =
\{\varphi_{\vec{\lambda}}\,|\,\vec{\lambda}\in
\RE^{\NN}\}$ is chosen in such a way they constitute an N-parameter
group of diffeomorphisms of $\cM$:
\begin{equation}
\begin{array}{cccc}
\varphi:&\cM\times\RE^{\NN} & \longrightarrow & \cM\\
&(p,\vec{\lambda})&|\!\!\!\longrightarrow&
\varphi(p,\vec{\lambda}):=\varphi^{}_{\vec{\lambda}}(p)\,. \\
\end{array}
\end{equation}
The identity element corresponds to $\vec{\lambda}=\vec{0}$,
i.e. $\varphi^{}_{\vec{0}}(p)=p$.  Moreover, a consistent perturbation
scheme should have the property that perturbing first with respect to a given
parameter, say $\lambda_{\PP}$, and afterwards with respect to another
parameter,
say $\lambda_{\QQ}$, must be equivalent to the converse operation.
We can implement this idea by imposing the following composition rule
for the group ${\cal G}[\varphi]$:
\begin{equation}
\forall\;\vec{\lambda}\,,\vec{\lambda'}\,,~~
\varphi_{\vec{\lambda}}\circ\varphi_{\vec{\lambda'}}=
\varphi_{\vec{\lambda}+\vec{\lambda'}}\,. \label{group}
\end{equation}
This property implies that the group is Abelian.  It also implies
that we can decompose $\varphi_{\vec{\lambda}}$ into N one-parameter
groups of diffeomorphisms (flows) that remain implicitly defined by the
equalities (we have N! equalities)
\begin{eqnarray}
\varphi_{\vec{\lambda}} & = &
\varphi^{}_{(\lambda_1,0,\ldots,0)}\circ\varphi_{(0,\lambda_2,\ldots,0)}
\circ\cdots\circ\varphi^{}_{(0,0,\ldots,\lambda_{\NN})} \nn \\
& = & \varphi^{}_{(0,\lambda_2,\ldots,0)}\circ
      \varphi^{}_{(\lambda_1,0,\ldots,0)}\circ\cdots\circ
      \varphi^{}_{(0,0,\ldots,\lambda_{\NN})} = \cdots \,. \label{philambda}
\end{eqnarray}
The action of the flow $\varphi^{}_{(0,\ldots,\lambda_{\MM},\ldots,0)}$
is generated by a vector field, $\zeta^{}_{\MM}$ (M$=1,\ldots,$N), acting on
the tangent space of $\cM$, and the Lie derivative of an arbitrary tensor field
$T$ with
respect to $\zeta^{}_{\MM}$ is given by
\begin{eqnarray}
\pounds^{}_{\zeta^{}_{\MM}}T= \left[\frac{\partial
\varphi^{\ast}_{(0,\ldots,\lambda_{\MM},\ldots,0)}T}
{\partial\lambda_{\MM}} \right]_{\lambda_{\MM}=0}\,,
\end{eqnarray}
where the superscript ${}^*$ denotes the pull-back.  Since the group
is Abelian, the vector fields $\zeta^{}_{\MM}$ commute
\begin{equation}
[\zeta^{}_{\PP},\zeta^{}_{\QQ}]=0 ~~~(\mbox{P,Q}=1,\ldots,\mbox{N}) \,.
\label{commutation}
\end{equation}
The Taylor expansion around $\vec{\lambda}=\vec{0}$ of the pull-back
associated with the flow
$\varphi^{}_{(0,\ldots,\lambda_{\MM},\ldots,0)}$ is given by
\begin{eqnarray}
\varphi^{\ast}_{(0,\ldots,\lambda_{\MM},\ldots,0)}T
=\sum_{k=0}^{\infty}\frac{\lambda^k_{\MM}}{k!}
\pounds^k_{\zeta^{}_{\MM}}T \,. \label{flowex}
\end{eqnarray}
This expression can be written in a more compact way by using the formal
exponential notation
\begin{eqnarray}
\varphi^{\ast}_{(0,\ldots,\lambda_{\MM},\ldots,0)}T =
\exp{(\lambda_{\MM}\pounds_{\zeta^{}_{\MM}})}\, T =
\e^{\lambda_{\MM}\pounds_{\zeta^{}_{\MM}}}\;T\,,\label{expnot}
\end{eqnarray}
which provides a clear operational way for working with groups
of diffeomorphisms. {}From expressions (\ref{flowex}) and (\ref{philambda})
we can derive the Taylor expansion of the pull-back
$\varphi^{\ast}_{\vec{\lambda}}T$
\begin{equation}
\varphi^{\ast}_{\vec{\lambda}}T \sum_{k_1,\ldots,k_{\NN}=0}^{\infty}
\left(\prod_{\mbox{\PP}=1}^{\mbox{\NN}}
\frac{\lambda^{k^{}_{\PP}}_{\PP}}{k_{\PP}!}
\pounds^{k_{\PP}}_{\zeta^{}_{\PP}}\right)\; T\,.\label{Taylorgroup}
\end{equation}
Using the exponential notation and the commutation relations (\ref{commutation})
we can write it as follows:
\begin{eqnarray}
\varphi^{\ast}_{\vec{\lambda}}T & = & \left[\prod_{\mbox{\PP}=1}^{\mbox{\NN}}
\exp{\left(\lambda_{\PP}\pounds^{}_{\zeta^{}_{\PP}}\right)}\right] T =
\exp{\left( \sum_{\PP=1}^{\NN} \lambda_{\PP} \pounds^{}_{\zeta^{}_{\PP}}
\right)}\; T \,.
\end{eqnarray}
In a given gauge $\varphi$, the perturbation of an arbitrary
tensorial quantity $T$ is defined as
\begin{equation}
\Delta T^\varphi_{\vec{\lambda}} :=
\varphi^*_{\vec{\lambda}}T^{}_{\vec{\lambda}}-T^{}_{\vec{0}}\,.
\label{defperturbation}
\end{equation}
The first term on the right-hand side of (\ref{defperturbation}) can be
Taylor-expanded around $\vec{\lambda}=\vec{0}$ using (\ref{Taylorgroup})
to get
\begin{equation}
\Delta T^\varphi_{\vec{\lambda}}=
\sum_{k_1,\ldots,k_{\NN}=0}^{\infty}
\left(\prod_{\mbox{\PP}=1}^{\mbox{\NN}}
\frac{\lambda_{\PP}^{k_{\PP}}}{k_{\PP}!}\right)\;
\delta^{\vec{k}}_\varphi T-T^{}_{\vec{0}}\,,
\label{perturbationexpansion}
\end{equation}
where $\vec{k}=(k_1,\ldots,k_{\NN})$ and
\begin{equation}
\delta^{\vec{k}}_\varphi T:=
\left[\frac{\partial^{k_1+\cdots+k_{\NN}}}
{\partial\lambda_1^{k_1} \cdots \partial\lambda_{\NN}^{k_{\NN}}}
\varphi^{\ast}_{\vec{\lambda}}T\right]_{\lo}
 = \prod_{\mbox{\PP}=1}^{\mbox{\NN}} \pounds^{k_{\PP}}_{\zeta^{}_{\PP}}\; T
\,, \label{defperturbationk}
\end{equation}
which defines the perturbation of order $(k_1,\ldots,k_{\NN})$ of
$T$ ($\delta^{\vec{0}}_\varphi T:=T_{\vec{0}}$).
The total order of a perturbation can be defined as $n_T:=k_1+\cdots+k_{\NN}$.
{}As a consequence of these definitions we have that
$\Delta T^\varphi_{\vec{\lambda}}$ and $\delta^{\vec{k}}_\varphi T$ are
fields that belong to the background spacetime model $M_{\vec{0}}$.

%\section{Gauge Transformations in N-Parameter Perturbation Theory}
%\label{sthree}
Let us consider two different gauges $\varphi$ and $\psi$, with generators
$(\vp\zeta^{}_1,\ldots,\vp\zeta^{}_{\NN})$ and
$(\ps\zeta^{}_1,\ldots,\ps\zeta^{}_{\NN})$ respectively.   For all $\vec{\lambda}$,
the objects defined in these two gauges can be related by a
diffeomorphism $\Phi^{}_{\vec{\lambda}}:\cM\rightarrow\cM$ given by
\begin{equation}
\Phi^{}_{\vec{\lambda}} :=
\varphi^{-1}_{\vec{\lambda}}\circ \psi^{}_{\vec{\lambda}}
= \varphi^{}_{-\vec{\lambda}}\circ\psi^{}_{\vec{\lambda}} \,.
\label{defGT}
\end{equation}
This is what is called a {\em gauge transformation} in perturbation
theory.  The family of all the possible gauge transformations for two given
gauges $\varphi$ and $\psi$
\begin{equation}
\begin{array}{cccc}
\Phi:&\cM\times\RE^{\NN} & \longrightarrow & \cM\\
&(p,\vec{\lambda})&|\!\!\!\longrightarrow&
\Phi(p,\vec{\lambda})=\Phi^{}_{\vec{\lambda}}(p)\,, \\
\end{array}
\end{equation}
is not in general a group of diffeomorphisms.
The action of the gauge transformation $\Phi_{\vec{\lambda}}$
can be written as
\begin{eqnarray}
\Phi^*_{\vec{\lambda}} T & = & (\varphi_{-\vec{\lambda}}
\circ\psi_{\vec{\lambda}})^\ast T =
\psi^\ast_{\vec{\lambda}}\circ\varphi^\ast_{-\vec{\lambda}}T
\nn \\
& = &  \exp{\left(\sum_{\PP=1}^{\NN}\lambda_{\PP}\pounds_{\ps\zeta^{}_{\PP}}\right)}
\exp{\left(-\sum_{\PP=1}^{\NN}\lambda_{\PP}\pounds_{\vp\zeta^{}_{\PP}}\right)}T \,.
\label{problem2}
\end{eqnarray}
Using group theory techniques we can write (\ref{problem2}) as the
action of a single exponential operator.  This can be explicitly
done by using the BCH formula \cite{BCH} (cf.\ \cite{MAB}).  This formula,
which can be applied to any two {\em linear operators} $A$ and $B$,
reads
\begin{eqnarray}
\e^A\e^B= \e^{f(A,B)} \,, \label{ch1}\\
f(A,B) = \sum_{m\geq 1}^{\infty} \frac{(-1)^{m-1}}{m}
\hspace{-5mm}
\sum_{\begin{array}{c} p_i,q_i \\
p_i+q_i\geq 1\end{array}}
\hspace{-5mm}
\frac{[\overbrace{A\cdots A}^{p_1}
\overbrace{B\cdots B}^{q_1} \cdots \overbrace{A\cdots A}^{p_m}
\overbrace{B\cdots B}^{q_m}]}{\left[\sum_{\alpha=1}^{m}(p_j+q_j)\right]
\prod_{\alpha=1}^{m}p_\alpha!q_\alpha!} \,, \label{ch2}
\end{eqnarray}
where the following notation has been used:
\begin{eqnarray}
[X_1X_2X_3\cdots X_n]:= [\cdots [[X_1,X_2],X_3],\cdots,X_n]\,. \label{ch3}
\end{eqnarray}
Then, the BCH formula can be seen as an expansion in commutators of the
initial operators $A$ and $B$.
Up to two commutators this expansion is given by
\begin{eqnarray}
f(A,B) = A + B + \frac{1}{2}[A,B] + \frac{1}{12}[[A,B],B] +
\frac{1}{12}[[B,A],A] + \cdots \,. \label{twocom}
\end{eqnarray}
This infinite expansion can be truncated and becomes
finite when some commutators vanish (for a solvable Lie algebra).
For instance, if $[[A,B],A] =[[A,B],B]=0$, then the BCH formula only
contains the three first terms shown in (\ref{twocom}).

The application of the BCH formula, (\ref{ch1}) and (\ref{ch2}),
to the construction
of the gauge transformation $\Phi_{\vec{\lambda}}$ (\ref{problem2})
constitutes the main point in our approach.   As we are going to see, it
provides an operational apparatus to compute all the perturbation orders
as well as expressions for the vector fields that generate the transformation
between different gauges.  This supposes an important advantage with respect
to the approach to two-parameter NLPT considered in \cite{BrGuSo},
which is based on a construction order by order and no close
expressions to every order can be obtained.
In what follows, we show how to use the BCH formula to obtain closed
expressions at every order,  in particular for the
gauge transformation generators.

To apply the BCH formula to Eq. (\ref{problem2}) we have to choose $A$ and $B$
in (\ref{ch1}) and (\ref{ch2}) as follows:
\begin{eqnarray}
A=\sum_{\PP=1}^{\NN}\lambda_{\PP}\pounds_{\ps\zeta^{}_{\PP}}
~~\mbox{and}~~
B=-\sum_{\PP=1}^{\NN}\lambda_{\PP}\pounds_{\vp\zeta^{}_{\PP}}\,. \label{abop}
\end{eqnarray}
Since $A$ and $B$ are linear in the parameters, the number of commutators in
a given term in the expansion of $f(A,B)$ coincides with the total
perturbation order $n_T$.

Using the properties of Lie derivatives we can then express the
gauge transformation $\Phi_{\vec{\lambda}}$ in the following way:
\begin{eqnarray}
\Phi^\ast_{\vec{\lambda}}T = \exp\left\{ \pounds^{}_{f\left(
\sum_{\PP=1}^{\NN}\lambda_{\PP}\ps\zeta^{}_{\PP},
-\sum_{\QQ=1}^{\NN}\lambda_{\QQ}\vp\zeta^{}_{\QQ}\right)}
\right\} T \,.
\end{eqnarray}
This can be rewritten as:
\begin{eqnarray}
\Phi^\ast_{\vec{\lambda}}T = \exp{\left\{\sum_{k_1,\ldots,k_{\NN}=0}^\infty
\left(\prod_{\PP=1}^{\NN}\frac{\lambda_{\PP}^{k_{\PP}}}{k_{\PP}!} \right)
\pounds_{\xi^{}_{\vec{k}}}-I\right\}}T \,, \label{exp2p}
\end{eqnarray}
where $\pounds_{\xi_{\vec{0}}}$ denotes the identity operator $I$ and
the rest of terms are Lie derivatives.  This is a consequence of the fact
that $A$ and $B$ are
linear combinations of Lie derivatives (\ref{abop}) and that the
functional $f(A,B)$ is a linear combination of $A$, $B$, and commutators
formed out of $A$ and $B$ (\ref{twocom}).  Then, this introduces
an infinite set of vector fields
$\{\xi_{\vec{k}}\,|\,\vec{k}\in\NA^{\NN}-\{\vec{0}\}\,\}$.
By direct comparison of (\ref{twocom},\ref{abop}) and (\ref{exp2p})
we can find the explicit expressions of these vector fields $\xi_{\vec{k}}$
directly in terms of the generators of the gauges $\varphi$ and $\psi$.

We can then derive an expression for the gauge transformation up to a given order
in the perturbation parameters by simply expanding the exponential (\ref{exp2p}).
Up to third total order ($n_T=3$) we obtain
\begin{eqnarray}
\Phi^\ast_{\vec{\lambda}}T & = & T + \sum_{\PP=1}^{\NN}\lambda^{}_{\PP}
\pounds^{}_{\xi^{}_{\vec{k}^{}_{\PP}}}T
+\frac{1}{2}\sum_{\PP,\QQ=1}^{\NN}\lambda^{}_{\PP}
\lambda^{}_{\QQ}\left(\pounds^{}_{\xi^{}_{\vec{k}^{}_{\PP}+\vec{k}^{}_{\QQ}}}+
\pounds^{}_{\xi^{}_{\vec{k}^{}_{\PP}}}\pounds^{}_{\xi^{}_{\vec{k}^{}_{\QQ}}}
\right)T \nn \\
& + & \frac{1}{6}\sum_{\PP,\QQ,\RR=1}^{\NN}\lambda^{}_{\PP}
\lambda^{}_{\QQ}\lambda^{}_{\RR}
\left( \pounds^{}_{\xi^{}_{\vec{k}^{}_{\PP}+\vec{k}^{}_{\QQ}+\vec{k}^{}_{\RR}}}
+\frac{3}{2}\left\{ \pounds^{}_{\xi^{}_{\vec{k}^{}_{\PP}+\vec{k}^{}_{\QQ}}},
\pounds^{}_{\xi^{}_{\vec{k}^{}_{\RR}}}\right\} \right. \nn \\
& + & \left. \pounds^{}_{\xi^{}_{\vec{k}^{}_{\PP}}}
\pounds^{}_{\xi^{}_{\vec{k}^{}_{\QQ}}}
\pounds^{}_{\xi^{}_{\vec{k}^{}_{\RR}}}\right)T + O^4(\vec{\lambda})\,,
\label{3order}
\end{eqnarray}
where $\vec{k}_{\PP} = (\overbrace{0,\ldots,0}^{\PP-1},1,
\overbrace{0,\ldots,0}^{\NN-\PP})$ and $\{A,B\}$ denotes the anticommutator
of $A$ and $B$.  {}From (\ref{3order}) we can easily derive the transformation
of a given perturbation from the gauge $\varphi$ to the gauge $\psi$,
$\delta^{\vec k}_{\psi}T-\delta^{\vec k}_{\varphi}T$.
First, from (\ref{defGT}), the pull-backs
$\varphi^*_{\vec\lambda}T$ and $\psi^*_{\vec\lambda}T$
are related by
\begin{equation}
\psi^*_{\vec\lambda}T_{\vec\lambda}=
\Phi^*_{\vec\lambda}
\varphi^*_{\vec\lambda}T_{\vec\lambda}\,.\label{phiPhipsi}
\end{equation}
Then, using (\ref{defperturbation}) and (\ref{perturbationexpansion}) we have
that
\begin{eqnarray}
\varphi^*_{\vec\lambda}T_{\vec\lambda}
&=&\sum_{k_1,\dots,k_{\NN}=0}^{\infty}
\left(\prod_{\PP=1}^{\NN}\frac{\lambda_p^{k_{\PP}}}{k_{\PP}!}\right)
\delta_{\varphi}^{\vec k}T\,, \label{expans}
\end{eqnarray}
and for $\psi^*_{\vec\lambda}T_{\vec\lambda}$ we only have to replace
$\varphi$ by $\psi$.
From (\ref{3order})-(\ref{expans}) we can extract the expressions for the
$\delta^{\vec k}_{\psi}T$'s in terms of Lie derivatives
of the $\delta^{\vec k}_{\varphi}T$'s.
 In the particular case N$=\!2$ and $\vec k\!=\!(1,1)$ we find
\begin{eqnarray}
\delta^{(1,1)}_{\psi}T&=&\delta^{(1,1)}_{\varphi}T+
\pounds_{\xi_{(1,0)}}\delta_{\varphi}^{(0,1)}T+
\pounds_{\xi_{(0,1)}}\delta_{\varphi}^{(1,0)}T\nonumber\\
&&+\left[\pounds_{\xi_{(1,1)}}+\frac{1}{2}\left\{
\pounds_{\xi_{(1,0)}},\pounds_{\xi_{(0,1)}}\right\}\right]T_0\,. \label{practical}
\end{eqnarray}

%\section{Standard Perturbation Theory: the N$=1$ Case}\label{sfour}
With the aim of comparing the formulation here introduced with previous
approaches to NLPT, we will show now how to recover standard
one-parameter NLPT. Let us consider two gauge choices: $\varphi$ and $\psi$.
For a given $\lambda$, the action of their associated pull-backs can be
written in the exponential notation as:
$\varphi^{\ast}_{\lambda}T=\e^{\lambda\pounds^{}_{\vp\zeta}}T$,
and $\psi^{\ast}_{\lambda}T=\e^{\lambda\pounds_{\ps\zeta}}T$.
A gauge transformation between these two gauges is then described
by $\Phi_{\lambda} = \varphi_{\lambda}^{-1}\circ\psi^{}_{\lambda}$.
Using the exponential notation, its action can expressed as follows
\begin{equation}
\Phi^{\ast}_{\lambda}T=\e^{\lambda\pounds^{}_{\ps\zeta}}
\e^{-\lambda\pounds^{}_{\vp\zeta}}T\,.
\label{GT1}
\end{equation}
The result of using the BCH formula can be written as
\begin{equation}
\Phi^{\ast}_{\lambda}T=\exp\left(\sum_{n=1}^{\infty}
\frac{\lambda^n}{n!}\pounds^{}_{\xi^{}_n}\right)T\,,
\label{constrnew}
\end{equation}
where $\{\xi_n~|~n \in \NA-\{0\}\}$ is a set of
generators of $\Phi$.  These can be expressed
 in terms of the gauge generators
$\vp\zeta$ and $\ps\zeta$.  From (\ref{twocom}) the three first generators are:
\begin{equation}
\xi_1 = \ps\zeta-\vp\zeta \,,~
\xi_2 = [\vp\zeta,\ps\zeta]\,,~
\xi_3 = \frac{1}{2}[\vp\zeta+\ps\zeta,[\vp\zeta,\ps\zeta]]\,.
\end{equation}
Up to third order,  (\ref{constrnew}) gives [the N=1
subcase of (\ref{3order})] :
\begin{eqnarray}
\Phi^\ast_{\lambda} T & = & T + \lambda\pounds^{}_{\xi_1}T +
\frac{\lambda^2}{2}\left(\pounds^{}_{\xi_2}+\pounds^2_{\xi_1}\right)T \nn \\
& + & \frac{\lambda^3}{3!}\left(\pounds^{}_{\xi_3}+
\frac{3}{2}\left\{\pounds^{}_{\xi_1},\pounds^{}_{\xi_2}\right\}+
\pounds^3_{\xi_1} \right)T +O^4(\lambda) \,. \label{newrec}
\end{eqnarray}
This form of $\Phi^\ast_{\lambda} T$ , derived through the BCH
approach, is not the same as the one obtained using KDs in~\cite{BrMaMoSo,SoBr}.
However, as we are now going to show, the resulting gauge transformations
are - order by order - equivalent, as expected, since  both cases are
expansions of the same exact expression (\ref{GT1}).
The KD is defined as \cite{BrMaMoSo,SoBr}:
\begin{equation}
\Phi^{(k)}_{\lambda}=  \phi^{(k)}_{\lambda^k/k!}\circ\cdots
\circ\phi^{(2)}_{\lambda^2/2}\circ\phi^{(1)}_{\lambda}\,, \label{kds}
\end{equation}
where the $\phi^{(n)}$ are one-parameter groups of diffeomorphisms (flows).
The main idea behind this concept is that a family of diffeomorphisms
$\{\Phi_{\lambda}~|~\lambda\in\RE\}$ can be approximated at a given
order in $\lambda$, say $k$, by a KD of order $k$.
Therefore one can approximate $\Phi_\lambda$ by $\Phi^{(k)}_{\lambda}$ in
the following sense \cite{SoBr}:
\begin{equation}
\Phi^\ast_\lambda T - \Phi^{(k)\ast}_\lambda T = O^{k+1}(\lambda)\,.
\end{equation}
The action of the pull-back of $\Phi^{(k)}_{\lambda}$ can be expressed,
using the exponential notation, as
\begin{equation}
\Phi^{(k)\ast}_{\lambda}T= \e^{\lambda\pounds^{}_{\chi^{}_1}} \cdots
\e^{\frac{\lambda^k}{k!}\pounds^{}_{\chi^{}_k}} T
\,, \label{constrold}
\end{equation}
where $\{\chi_n\}^{}_{n=1,\dots,k}$ is the set of generators of the
family $\Phi^{(k)}$, and  each $\chi_n$ is the
generator of the flow $\phi^{(n)}$.  Like the $\xi_n$'s,
they can be expressed in terms of the gauge generators
$\vp\zeta$ and $\ps\zeta$. Hence, we can   find the relations between
the  $\xi_n$'s and $\chi_n$'s. Up to third order we have
\begin{eqnarray}
\xi_1 = \chi_1\,,~~
\xi_2 = \chi_2\,,~~
\xi_3 = \chi_3+\frac{3}{2}[\chi_1,\chi_2]\,.
\label{chicazzo}
\end{eqnarray}
Therefore, the expansion for $\Phi^{\ast}_\lambda T$ that we obtain from
the expansion of $\Phi^{(k)\ast}_\lambda T$ is (up to third order):
\begin{eqnarray}
\Phi^{(k)\ast}_{\lambda} T & = & T + \lambda\pounds^{}_{\chi^{}_1}T +
\frac{\lambda^2}{2}\left(\pounds^{}_{\chi^{}_2}+\pounds^2_{\chi^{}_1}\right)T
\nn \\
& + & \frac{\lambda^3}{3!}\left(\pounds^{}_{\chi^{}_3}+3\pounds^{}_{\chi^{}_1}
\pounds^{}_{\chi^{}_2}+\pounds^3_{\chi^{}_1} \right)T + O^4(\lambda)
\,, \label{3knight}
\end{eqnarray}
i.e.\ the result in \cite{BrMaMoSo,SoBr}.
Comparing the expansions (\ref{newrec}) and (\ref{3knight}) we see
that they have different structures. However, substituting $\chi_1$,
$\chi_2$ and $\chi_3$ from (\ref{chicazzo}) into (\ref{3knight}) we
obtain (\ref{newrec}); thus, these two expansions of the gauge
transformation (\ref{GT1}) are equivalent.  Our formulation,
Eq. (\ref{newrec}), leads to an expansion with
terms of the form $\cdots \pounds_{\xi^{}_k}\cdots
\pounds_{\xi^{}_l}\cdots T$ with $l<k\,,$ which do not appear
in the formulations of \cite{BrMaMoSo}
[due to the ordering introduced by the KDs, see (\ref{kds})] and \cite{BrGuSo}.
Comparing further our formulation with the order by order approach
in \cite{BrGuSo} we
see how the use of the BCH formula naturally selects a unique
expansion for the gauge transformation, in contrast with \cite{BrGuSo},
which contains freely specifiable parameters.

To sum up, we have presented a formulation of N-parameter NLPT  in which
we have a (unique) closed formula for the expansion of general gauge
transformations and whose consistency is given by construction, shedding
light onto the underlying mathematical structure.  The importance of this
result is even more clear if considered in the context of practical
applications of relativistic perturbation theory, where the issue of
comparing results obtained in different gauges and the related problem of
constructing gauge-invariant formulations have always been crucial to obtain
physically transparent results \cite{Sa,BrSo}.  These issues become even
more important when dealing with non-linear perturbations \cite{BrMaMoSo,SoBr,BrSo}
and more than one parameter. Our formalism provides
the  gauge transformations and the conditions for gauge invariance for every
perturbation order in N-parameters
(explicit conditions for the 2-parameter case where given in \cite{BrGuSo}).

In retrospect, one may wonder why our general results have not been previously
derived, given that the BCH formula  has long been known in differential
geometry \cite{BCH}. The  likely answer is that, although the problem of gauge
dependence is as old as relativistic perturbation itself \cite{Sa}, untill
recently spacetime perturbations have mostly been considered at first order
only and for a single parameter, and consequently gauge transformations have
always been dealt with at the most elementary linear level, where the
BCH formalism and the exponentiation (\ref{expnot}) on which it is based
are superfluous.  When the problem of gauge transformations has been
considered in NLPT for the case of one parameter, two routes have been
followed. In \cite{BrMaMoSo,SoBr} KDs have been introduced and used (see
also \cite{FlWa} for an equivalent second order result and \cite{bi:GL} for
some basic fomulas), and in particular in \cite{SoBr} a closed formula was
derived to generate gauge transformations at arbitrary order. In \cite{MAB}
on the other hand the BCH formula was used, for one parameter at second
order. As we have illustrated  above in the one parameter case and up to
third order, the two routes are equivalent in that they provide equivalent
gauge transformations at the required order.
On the other hand, the gauge transformations derived in \cite{BrGuSo} contain
freely specifiable constants (linked by sets of constraints) that are not
present in the BCH derivation presented here.
Again, order by order the gauge transformations are equivalent, with one
specific choice of the constants corresponding directly to the BCH derivation,
and other choices connected by appropriate relations between the two sets
of generators of the gauge transformations.
From the point of view of generality elegance and compactness of the
derivation the BCH approach presented here is by far superior to that
followed in \cite{BrGuSo}. However, for practical purposes one is interested
in the gauge transformations at a given order, e.g.\ (\ref{practical}),
and in this perspective we believe that the   formulas  with freely
specifiable constants in \cite{BrGuSo} may still be useful.
Indeed, the typical problem  (see e.g.\ \cite{BrMaMoSo,MaMoBr}) is that
one wants to know how to transform between two pre-assigned gauges. In this
case the unknowns of the problem are the generators of the transformation.
Then, one faces integration calculations, and given that  two different
choices of the freely specifiable constants  correspond to integration in
a different order, it may turn out that one specific choice of  constants
is better in solving the problem.

We want to finish by discussing the potential applications of N-parameter NLPT.
First of all, it is important to remark that perturbation theory in general relativity,
and in other spacetime theories (some of them of
great  relevance nowadays),
remains the main alternative to fully numerical
methods in a context in which one has to deal with sets of non-linear field
equations.  Depending on the physical problem at hand, it is sometimes necessary
to go beyond simple linear perturbations, considering higher-order contributions.
In this sense, a multi-parameter perturbative scheme as the one presented here
allows us to select only the higher-order perturbative sectors relevant for the
physical problem under consideration, simplifying in this way the calculations involved.

There are already quite few applications of the one-parameter NLPT at second
order in the literature.  In cosmology, the evolution of perturbations in two different
gauges is explicitly worked out \cite{BrMaMoSo,MaMoBr} and applications to
the Cosmic Microwave Background have been considered~\cite{MoMa}. Further applications
in the cosmological context can be found in \cite{BrMeTa,MAB}.
In an  astrophysical context there are a number of studies  of  sources of
gravitational radiation:
in \cite{CPM,collapse} oscillations during gravitational collapse have been
analyzed; in \cite{closelimit}  the  so called close limit approximation is used to study
the outcome of  black hole mergers; in  \cite{CaLo}
a second-order gauge-invariant perturbative scheme for
the Kerr metric has been developed.
The N-parameter NLPT opens the door for new applications in spacetime
theories (see \cite{copione2} for an example).  In the general relativistic case,
it can provide a new way of
studying slowly rotating relativistic stars, and it can be the main tool
to study other issues as for instance the non-linear coupling of oscillation
modes of relativistic stars, or in cosmology to study the combined
effect of magnetic fields and linear perturbations in the matter distribution.

%\section*{Acknowledgement}
MB thanks the Dipartimento di Fisica ``G. Marconi'' (Universit\`a di Roma
``La Sapienza'') and LG the Institute of Cosmology and Gravitation
(Portsmouth University) for hospitality.
CFS is supported by the EPSRC. This work has
been partly supported by the European Commission (Research Training Network
HPRN-CT-2000-00137) and by INFN.

%%%%%%%%%%%%%%%%%%%%
% THE BIBLIOGRAPHY
%%%%%%%%%%%%%%%%%%%%

\end{document}